 \documentclass{vgtc}                          

\ifpdf
  \pdfoutput=1\relax                   
  \usepackage{graphicx}                
  \DeclareGraphicsExtensions{.pdf,.png,.jpg,.jpeg} 
\else
  \usepackage{graphicx}                
  \DeclareGraphicsExtensions{.eps}     
\fi%

\graphicspath{{figures/}{pictures/}{images/}{./}} 

\usepackage{microtype}                 
\PassOptionsToPackage{warn}{textcomp}  
\usepackage{textcomp}                  
\usepackage{mathptmx}                  
\usepackage{times}                     
\usepackage{cite}                      
\usepackage{tabu}                      
\usepackage{booktabs}                  

\usepackage{times}                     

\usepackage{lipsum}                    
\usepackage{mwe}                       

\usepackage[hyphens]{url}                       
\usepackage{ulem}                     

\usepackage{changes} 

\onlineid{0}

\vgtccategory{Research}

\title{ChatAR: Conversation Support using Large Language Model and Augmented Reality}

\author{Yuichiro Fujimoto\thanks{e-mail: yfujimoto@rins.ryukoku.ac.jp}\\ %
        \scriptsize Ryukoku University}

\teaser{
  \centering
  \includegraphics[width=1.0 \columnwidth]{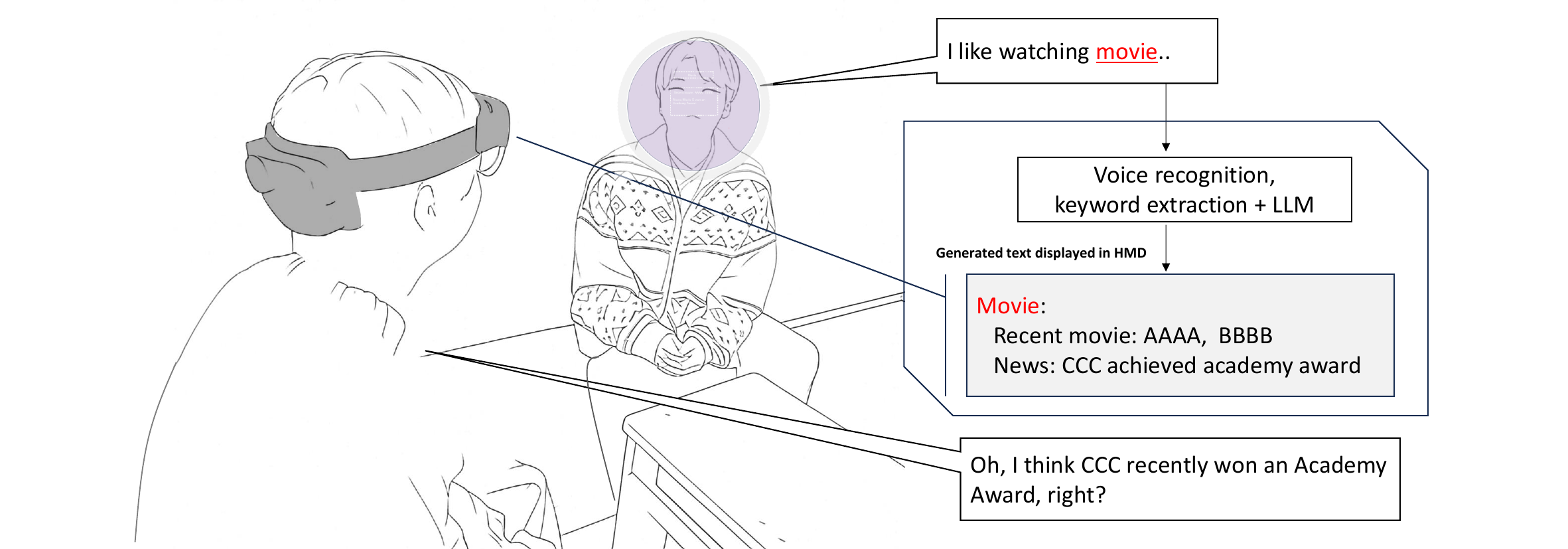}
  \caption{ChatAR – Conversation Support Using a Large Language Model and Augmented Reality.
ChatAR automatically detects keywords from conversational speech and generates relevant information in real time. This information is displayed on the HMD in a way that minimizes disruption to the user's eye movements while ensuring effective communication.
  }
  \label{fig:teaser}
}

\abstract{
Engaging in smooth conversations with others is a crucial social skill. However, differences in knowledge between conversation participants can sometimes hinder effective communication.
To tackle this issue, this study proposes a real-time support system that integrates head-mounted display (HMD)-based augmented reality (AR) technology with large language models (LLMs).
This system facilitates conversation by recognizing keywords during dialogue, generating relevant information using the LLM, reformatting it, and presenting it to the user via the HMD.
A significant issue with this system is that the user's eye movements may reveal to the conversation partner that they are reading the displayed text.
This study also proposes a method for presenting information that takes into account appropriate eye movements during conversation.
Two experiments were conducted to evaluate the effectiveness of the proposed system.
The first experiment revealed that the proposed information presentation method reduces the likelihood of the conversation partner noticing that the user is reading the displayed text.
The second experiment demonstrated that the proposed method led to a more balanced speech ratio between the user and the conversation partner, as well as a  increase in the perceived excitement of the conversation.
}


\begin{document}
 \maketitle

\section{Introduction}
Face-to-face conversation forms the foundation of social life. 
Through conversations with others, we not only share information but also build relationships and contribute to the smooth functioning of society.
In particular, casual conversations without a specific purpose often play a crucial role in building relationships \cite{Garfinkel1967}.
This is attributed to human characteristics, such as the tendency to build trusting relationships and rapport more easily by sharing personal information and understanding others' values \cite{Laurenceau1998}.
However, many individuals feel uncomfortable or even fearful when engaging in casual conversations with others. 
It is estimated that over one million people in Japan experience social withdrawal due to this issue \cite{Fujimoto2020}.
Differences in the ability of both parties to engage in face-to-face conversations, gaps in background knowledge, and a lack of consideration for the other’s perspective can hinder the maintenance of smooth conversations \cite{Garfinkel1967, Krauss1991}.
This study focuses on addressing the issue of knowledge gaps between two individuals.

In conversations, it is essential to comprehend the other person's speech, integrate it with one's own knowledge, and respond appropriately.
If the response evokes empathy or interest in the other person, the conversation is likely to progress smoothly and positively impact the relationship.
On the other hand, if the responses are consistently lacking in substance, the conversation partner may develop a somewhat negative impression.
For example, consider a scenario where John and Adam, who have just met at a buffet party, are engaged in conversation.
Suppose John initiates a conversation with Adam by saying, ``My hobby is watching movies."
If Adam responds by saying, ``Oh, I don't watch movies, so I don't know anything about them," the conversation is likely to end there.
John, having disclosed personal information (i.e., his hobby), may feel somewhat disappointed and find it challenging to continue the conversation with Adam.
Even without any knowledge of movies, there are several ways to expand the conversation; however, this requires conversational skills.
Moreover, having broader general knowledge is often more advantageous than having limited knowledge.
Conventional general and medical approaches to addressing this issue typically involve conversation practice or social skills training, both of which require a long time to achieve improvement.

To address this issue, this study proposes ``chatAR," a chat support system that provides continuous intervention during face-to-face conversations by leveraging augmented reality (AR) and large language model (LLM) technologies, which have advanced remarkably in recent years.
LLM-based dialogue technology, including OpenAI's chatGPT \cite{chatGPT} and Google's Gemini \cite{Gemini}, has achieved smooth conversations between people and systems at a truly practical level.
Regarding the combined use of AR and (multimodal) LLMs, research on authoring tool applications has recently emerged (e.g., \cite{Chen2025, Srinidhi2024}). However, their application in the context of conversation support remains limited.
By recognizing the topic of a person-to-person conversation and inputting the extracted keywords into LLM-based dialogue technology, it becomes possible to generate relevant information and appropriate responses.
Furthermore, by integrating AR technology with a head-mounted display (HMD), the generated responses can be presented at an appropriate position within the user's field of view.
As of 2025, wearing an AR-HMD on a daily basis is not yet common. However, devices such as the Xreal One \cite{XrealOne} and Meta Orion \cite{MetaOrion}, which offer improved comfort for everyday use, are beginning to emerge.
The proposed system is designed with the assumption that, in the near future, lighter AR-HMDs will replace smartphones and be used continuously in daily life, including during face-to-face conversations.

On the other hand, the use of such a system may raise concerns that LLM will significantly intervene in the content of conversations, which could potentially change the meaning of human conversations and have a negative impact on users.
Therefore, this study takes these concerns into consideration and limits LLM intervention to a minimum, by occasionally providing users with conversation options.

Another important consideration when using the proposed method is that ``the system must not hinder smooth conversation."
For example, if the text generated by ChatGPT is simply displayed horizontally in front of the user via the HMD, the user would need to move their eyes repeatedly from side to side to read it, which could be disruptive.
The unnatural movement of the user's eyes may immediately make the conversational partner aware that the user is referring to something. 
While this is not necessarily undesirable, behaviors that clearly indicate a lack of focus on the conversation should be avoided \cite{Dwyer2018}.
Similarly, if the system requires user operation, it is preferable that the conversation partner remains unaware of this interaction. 
This study also proposes an information presentation method designed to make it difficult for the conversational partner to detect when the user is checking displayed information, which is incorporated into the LLM-AR conversation support system.
Furthermore, two experiments were conducted to evaluate both the effectiveness of the proposed information presentation method and the overall conversation support system.

To summarize, the contributions of this study are as follows:

\begin{itemize}
  \item This study proposes a conversation support framework that fully integrates LLM and AR technologies, along with a prototype implementation. This framework has the potential to be utilized in everyday conversations in the near future.
  \item An information presentation method is introduced that minimizes the likelihood of the conversational partner noticing when the user is referencing displayed information within the conversation support system.
  \item Two experiments demonstrated that the proposed system enhances the subjective excitement of conversations while reducing the conversational partner’s awareness of the user’s information references.
\end{itemize}

\section{Related Work}
This study focuses on conversation support systems. 
First, we review previous studies on conversation support using LLMs without AR and those utilizing AR separately. 
Next, we examine research on information presentation methods designed to minimize interruptions in conversations and other human activities.

\subsection{Conversation Support Using LLM}
In recent years, LLMs have been utilized in various professional contexts.
Wu et al. proposed AutoGen, a framework that enables flexible definitions of human-agent and agent-agent interactions through multiple agents \cite{Wu2023}.
One application of this framework is in mathematical problem-solving, where students and math experts communicate via an assistant agent.
Similarly, Yang et al. proposed Talk2Care, an asynchronous LLM-based communication system that facilitates interactions between home-based elderly individuals and medical professionals \cite{Yang2024}.
Hua et al. proposed an LLM-based agent that facilitates dialogue in business negotiations \cite{Hua2024}.
In contrast to these studies, which focus on conversations with specific and explicit goals, this study primarily aims to support ``small talk" in everyday conversations.

\subsection{AR-based Face-to-face Conversation Support System}
Xu et al.'s SocioGlass recognizes the face of a conversation partner during face-to-face communication and retrieves relevant information about that person from a database, displaying it to the user via Google Glass \cite{Xu2016}.
Similarly, Rivu et al. proposed a method that utilizes eye trackers to present information with minimal disruption to the conversation \cite{Rivu2020}.
The objectives and challenges addressed in these studies are similar to those of our study.
In the proposed method, information is displayed around the head of the conversation partner to reduce cognitive load. 
Jadon et al. proposed RealityChat, a HoloLens-based interface that leverages real-time speech recognition, natural language processing, and gaze-based interaction to instantly embed visual references \cite{Jadon2024}.
While this system shares technological similarities with our proposed method, its focus differs in terms of usage.
Specifically, RealityChat explores a novel form of communication in which information is sometimes shared visually during the conversation.
SocialMind, proposed by Yang et al., is a flexible system designed to adapt to various conversational situations \cite{Yang2024}.
It infers the persona of the conversation partner and suggests socially appropriate speech and behavior to the user by analyzing not only audio but also non-verbal cues and the partner's distance, based on image data from a camera.
However, considerations regarding the potential inhibition of conversation and strategies for appropriate information presentation methods are beyond the scope of that study.

In addition to displaying information that serves as a cue for directing conversation content, various forms of AR-based conversation support have also been explored.
Damian et al. proposed a system that evaluates a user's speech volume and gestures during conversation or public speaking and provides real-time feedback via smart glasses \cite{Damian2015}.
Yoneyama et al. introduced a method aimed at individuals with social anxiety tendencies, in which the original facial expressions and eye movements of the conversational partner are replaced with an AR avatar to reduce psychological pressure \cite{Yoneyama2024}.
Valente et al. proposed an AR-based approach that visualizes the emotions of a conversation partner, estimated using an emotion recognition model (AuRea) trained on ECG physiological data, to facilitate more effective empathic communication \cite{Valente2022}.

\subsection{AR Information Presentation Without Disturbing User}
The extent to which the presentation of information during conversation and other activities interferes with these interactions has been widely discussed in the literature (e.g., \cite{Dwyer2018}).
Ofek et al. examined the degree to which the amount and timing of text information displayed on a half-mirrored stationary device disrupted dialogue \cite{Ofek2013}.
Plabst et al. explored the effects of four different notification locations in an AR environment—subtitles, head-up, world space, and the user’s wrist—on user attention and task performance \cite{Plabst2022}.
Pfeuffer et al. introduced ``ARtention," a design space for gaze-based interaction in AR information interfaces \cite{Pfeuffer2021}.
This framework considers transitions from reality to virtual interfaces, from single-layer to multi-layered content, and from passive information consumption to interactive selection tasks.
Cai et al. proposed a semi-transparent circular menu for HMDs, called ParaGlassMenu, designed to mitigate conversational interruptions caused by digital device operations in social situations, including face-to-face conversations \cite{Cai2023}.  
The system displays menu items around the conversation partner’s face, allowing users to operate it discreetly via a ring-shaped device.

\subsection{Position of This Work}
This study proposes a method for supporting daily conversations by integrating LLM and AR technologies.
As discussed above, several similar studies have emerged in recent years within this context.
However, this study distinguishes itself by focusing on two key aspects: (1) enhancing the quality of casual conversations and (2) minimizing disruptions for both the user and the conversational partner.

\section{Ideas}
\label{Ideas}

\subsection{How LLM should be used?}
\label{How LLM should be used?}
Recent advancements in LLMs (generative AI) have significantly improved the ability to generate information tailored to specific requirements.
Technically, an LLM could generate all responses, synthesize speech with a tone modulated to match the user's voice, and present it to conversation partners \cite{Zhao2023, Sandler2024}.
However, fully automating responses without incorporating the user's intentions poses the risk of losing essential aspects of natural conversation.
In addition to the above, it is important to design systems that reduce dependence on LLM usage.
It is generally known that excessive support can lead to redundant dependence.

Therefore, in this study, the use of LLMs is restricted to supporting the user's actions to a minimum rather than fully automating responses.
Specifically, LLMs provide visual cues for conversation, allowing the user to decide whether and how to incorporate them into speech. 
The user may choose to speak the cue as presented, modify it, or generate a new response inspired by it.
Displaying excessive information can divert the user's attention for an extended period, impairing judgment and potentially disrupting the conversation \cite{Roetzel2019}.
Conversely, insufficient information reduces the likelihood of discovering useful conversational hints.
Thus, it is crucial to generate information that strikes a balance between these two factors.

\subsection{Overview of System}
\label{Overview of System}
This section provides an overview of the proposed system, ChatAR.
Since eye contact is a crucial social cue in conversations, an optical see-through (OST) head-mounted display (HMD) with a transparent front surface was selected as the display device.
Although continuous use of OST-HMDs in daily life remains uncommon as of 2025, this study is based on the assumption that the widespread adoption of smaller, more discreet OST-HMDs will become socially accepted in the near future.
Consider a scenario in which the user is engaged in a face-to-face conversation. 
ChatAR automatically detects keywords spoken by either the user or the conversation partner. 
If the user wishes, the system utilizes an LLM to generate relevant information based on the detected keywords.
 This information is then presented via AR within the user’s field of view in a manner that minimizes disruption to the conversation.
Through this process, ChatAR subtly supports the user's ability to maintain smooth and engaging conversations.

\subsection{System-applicable conversation}
\label{}
As mentioned above, since the purpose of this study is to eliminate disparities in general knowledge between conversation partners, the system's basic operation is to present information related to the keywords it recognizes.
In other words, effective conversation is limited to such dialogue situations.
For example, topics related to individuals who are not famous, abstract topics, and situations that require specific personal opinions are not covered in this study.

\subsection{Information Presentation Method}
\label{Information Presentation Method}
The fundamental requirement of the proposed system is to support idea generation in conversation without interfering with the conversation. 
Here, interference with the conversation encompasses two aspects: (a) interference with the user and (b) interference with the conversational partner.
While these two factors are interrelated, they are discussed separately in this study for clarity.

The primary modalities for conveying large amounts of textual information to users in real time are visual and auditory stimuli.
Since listening to the conversational partner is essential during a conversation, auditory stimuli take precedence.
Consequently, this study focuses on the use of visual stimuli for information presentation.

\subsubsection{Interference with the user}
\label{Interference with the user}

``(a) Interference with the user" refers to situations where the displayed information unnecessarily diverts the user's attention, disrupts their train of thought, and hinders smooth conversation.
To prevent this, information presentation should be carefully designed to avoid disturbing the user's cognitive flow or real-time interaction.
First, the intensity (e.g., brightness) of the visual stimulus should be set to the lowest level that ensures clear visibility.
Additionally, research has shown that information appearing in unexpected locations can significantly distract the user’s attention (e.g., \cite{Fujimoto2012}).
Therefore, even if the location of the displayed information changes, it should remain predictable for the user.
Furthermore, previous studies have highlighted that presenting excessive amounts of information can be undesirable for users (e.g., \cite{Rivu2020}).
Thus, the system should limit the displayed content to an amount that can be comprehended within a short period.

\subsubsection{Interference with the conversational partner}
\label{Interference with the conversational partner}

``(b) Interference with the conversational partner" refers to instances where the system’s information presentation unnecessarily diverts the attention of the conversation partner.
 This includes situations where the partner notices that the user’s HMD is displaying information or realizes that the user is referring to the displayed content.
The former depends on the display mechanism of the HMD.
For example, in HoloLens 2, presented information can be visible to surrounding individuals due to reflected light.
If the display luminance is high, the presence of displayed information becomes more noticeable.
Regarding the latter, if the user's gaze moves repeatedly in a predictable pattern (e.g., cyclically from left to right), the conversation partner may recognize that the user is reading text.
Such behavior, which clearly indicates a lack of conversational focus, should be avoided \cite{Dwyer2018}.
Therefore, in the proposed system, it is crucial to ensure that information is presented in a manner that minimizes the likelihood of detection by the conversational partner.

\subsubsection{Gaze during conversation}
\label{Gaze during conversation}
The topic of gaze behavior during dialogue has long been studied in the fields of communication studies and sociology \cite{Cook1977, Duncan1979, Stack1982, Hessels2020}.
In general, it is considered desirable to direct one’s gaze within the inverted triangle formed by the conversation partner’s eyes and nose.
Cook investigated the duration and timing of mutual gaze in face-to-face conversations \cite{Cook1977}.
The findings suggest that prolonged eye contact often causes discomfort for many individuals. 
Conversely, shorter gazing durations have been associated with signs of anxiety \cite{Waxer1977}.

In addition, it is generally recognized that large lateral eye movements are undesirable during conversation.
When breaking eye contact, directing one's gaze downward is considered the most natural approach \cite{Kobuki2010}.

Considering these factors, we decided to display text information within the boundaries of the inverted triangle formed by the conversational partner’s eyes and nose.
To prevent excessive eye contact, the system is designed to gradually shift the displayed information from the top to the bottom at regular intervals.
After a certain period, the text returns to its original position, facilitating more natural eye movement for the user.

\section{Implementation}
\label{Implementation}
In the current prototype, we utilized Microsoft HoloLens 2, an OST-HMD that allows the user’s eye movements to remain visible to the conversational partner and is equipped with an eye-tracking system.

\subsection{Functional Module}
\begin{figure}[t]
 \centering
  \includegraphics[width=1.0\columnwidth]{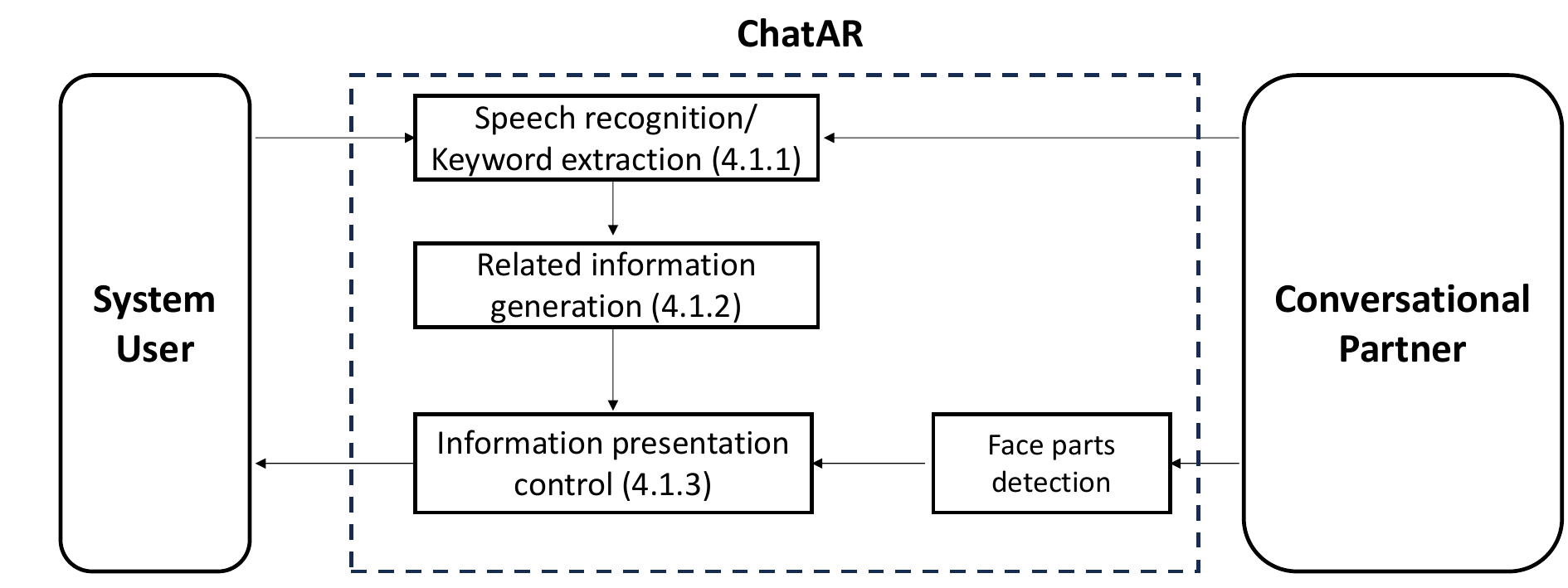}
  \caption{Processing configuration within the proposal system}
 \label{fig:module-overview}
\end{figure}

The ChatAR framework is composed of three primary modules: (1) the speech recognition module, (2) the related information generation module, and (3) the information presentation control module.
An overview of the system architecture is illustrated in Figure \ref{fig:module-overview}.

\subsubsection{Speech Recognition Module}
\label{KeywordReconigitionModule}

For speech recognition, we utilized the ``Speech to Text" function of the Microsoft Azure Speech Services API \cite{SpeechServices}.
The audio data captured by the HoloLens 2 microphone is transmitted as input to the SpeechRecognizer module.
The recognized text is then asynchronously forwarded to the subsequent related information generation module via the ChatCompletions module.
The system recognizes only the user's voice based on the microphone position and volume settings of the HMD.
The process of extracting prominent keywords from user speech is performed automatically by the prompt in the next section.

\subsubsection{Related Information Generation Module}
\label{GenerationModule}

We utilized Azure's OpenAI API \cite{OpenAI} and GPT-4o as the model to generate relevant information for the recognized keywords.
The agent was pre-assigned a specific role outlining its usage conditions and output restrictions.
``You are an AI that supports users in having smooth conversations with others. First, extract distinctive keywords from the input text. Next, Please provide a concise list of recent news and topics related to the keywords you have extracted. Please keep each response to 130 characters or less. Please avoid overly general content. Also, if there is no new input, please do not delete it until 300 seconds have passed since the previous response."
It is important to note that the actual prompt was written in Japanese, and the description above is an English translation.
Fine-tuning was not conducted, as preliminary trials demonstrated that the generated information met our accuracy requirements sufficiently.

\subsubsection{Information Presentation Control Module}
\label{InformationControlModule}

Based on the discussion in Sections 3 and \ref{GenerationModule}, the amount of text displayed at one time is limited to 130 characters.
To maintain natural eye contact, the text is primarily displayed in the area corresponding to the face of the conversational partner.
We use Dlib FaceLandmarkDetector, a library for detecting the 3D position and posture of the face and each part of the face \cite{Dlib}.
The width of the text display area was set to 90.0 mm, which corresponds to the average distance between the outer corners of the eyes of Japanese adults (our target user group), ensuring that the text remains within the inverted triangle connecting the eyes and nose \cite{AISTFace}. 
Similarly, the height was set to 50.0 mm, the average distance from the midpoint between the eyes to the bottom of the nose.
Although these measurements slightly differ between men and women, we adopted the smaller values (based on female averages) to ensure that the display remains within the appropriate region.
Additionally, to mitigate the discomfort associated with constant visual focus on the face area, as well as to align with general eye contact dynamics and natural gaze aversion behavior (see Section \ref{Gaze during conversation}), the text display automatically shifts downward by approximately 10 cm after five seconds of user gaze fixation on the conversational partner's face.
The text then returns to its original position after three seconds, allowing for more natural eye movement.

\subsection{User Interfaces}
\label{UserInterface}

\begin{figure}[t]
 \centering
  \includegraphics[width=\columnwidth]{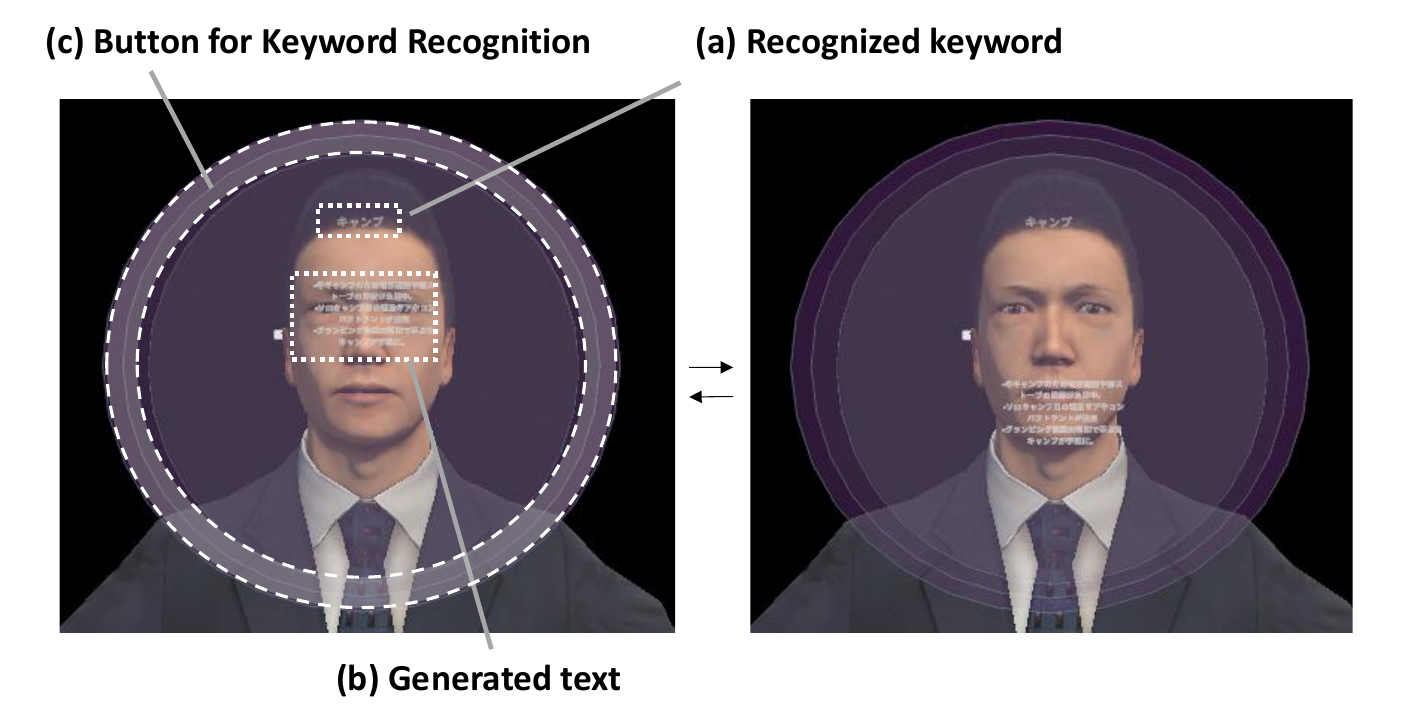}
  \caption{User interface of the proposed system. (For better readability in print, the text is displayed with increased brightness. The character shown is from Microsoft RocketBox \cite{Gonzalez-Franco2020}.)}
 \label{fig:UI}
\end{figure}

From the discussion in Section \ref{Ideas}, the primary goal of the user interface design for the proposed system was to ensure minimal interference with the conversation (i.e., avoiding disruption for both the user and the conversational partner). 
Figure \ref{fig:UI} illustrates the user interface of the proposed system.
All elements are placed within a circular panel to help users maintain an overview of the displayed information without distraction.
A dark purple background is used to enhance text visibility. 
The recognized keywords (a) are displayed at the top of the circle, while the generated text (b) appears in the center. 
As described in Section \ref{InformationControlModule}, this hint text automatically moves up or down over time to support natural gaze behavior.

When testing the initial prototype, we found that the system sometimes recognized keywords when it was not necessary.
To address this, we added a button (Figure \ref{fig:UI} (c)) that allows users to manually toggle keyword recognition on and off.
If the user gazes at one of the button areas for more than one second, the system will switch keyword recognition and hint text generation on or off.
To ensure ease of use without disrupting natural eye movement during conversation, buttons were placed around the entire circumference of the circular panel rather than in a single fixed location.

\subsection{Examples of Assistance During Chat}
\begin{figure}[t]
 \centering
  \includegraphics[width=1.0 \columnwidth]{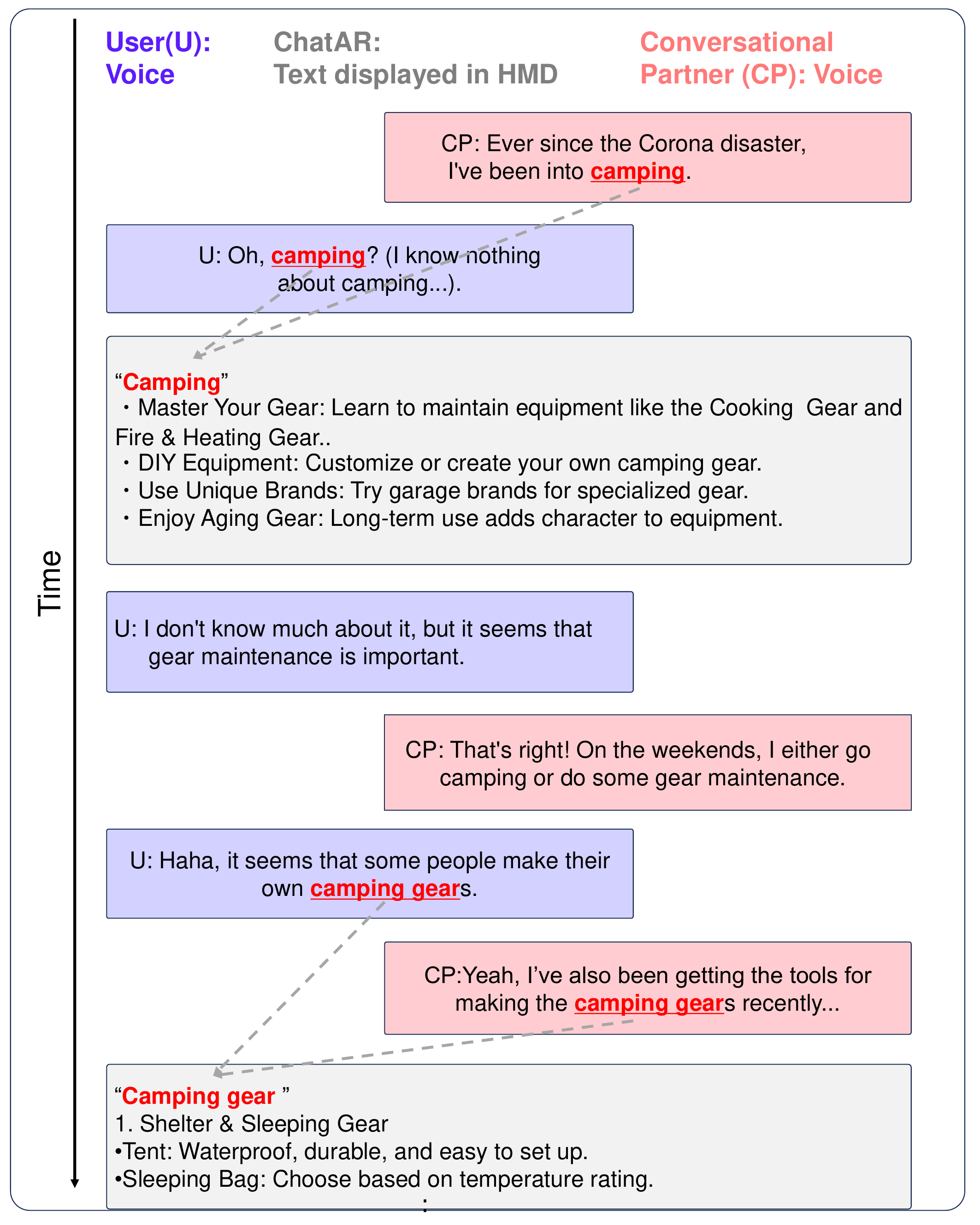}
  \caption{Examples of assistance during chat.}
 \label{fig:conversation-example}
\end{figure}

Figure \ref{fig:conversation-example} illustrates an example of conversation support using ChatAR.
First, the conversational partner mentions that camping is their latest hobby.
While the user is still formulating a response, he repeats, ``Oh, Camping?".
Here, the system detects the word ``camp" twice, triggering the generation of relevant information, which is then presented to the user in a bulleted list. 
Using this as a conversational cue, the user inquires about the maintenance of camping gear.
In response, the conversational partner shares that he has been engaged in camping gear maintenance throughout the weekend.
The user further expands the conversation to making one's own camping gear.
Detecting the repeated mention of ``camping gear" by both the user and the conversational partner, the system subsequently presents information related to ``camping gear," allowing the conversation to progress naturally.

As this example shows, the proposed system only provides conversational cues, from which the user can construct a response as he or she is comfortable speaking.
This system design allows the system to support smooth conversation without compromising the framework of human-to-human conversation (i.e., without changing it to AI-to-human).

\section{Experiment1}
\label{experiment1}

Experiment 1 was conducted to assess the impact of the proposed information presentation method on the user's eye movements.
Both Experiment 1 and Experiment 2 were approved by the Institutional Review Board (IRB) [Details removed for peer review].

\subsection{Overview}
In Experiment 1, we evaluated whether the conversational partner could perceive when the system user was viewing the displayed information using the proposed information presentation method.
 The experiment was conducted in pairs, with participants who had never met before.
One participant wore the HoloLens 2 and acted as the system user, while the other participant served as the conversational partner.
The participants sat facing each other under the assumption of engaging in a casual conversation.
The system user silently read the displayed text while attempting to minimize any noticeable indications of reading.
Meanwhile, the conversational partner was instructed to estimate and report the duration for which they believed the system user was reading something.
By comparing these results with those obtained using conventional information presentation methods, we assessed the effectiveness of the proposed approach.

\subsection{Participants}

Sixteen participants were recruited for this study from a specific department of a local university via an on-campus electronic bulletin board. All participants were undergraduate students in information science and native Japanese speakers (14 males, 2 females), aged 20–25 years ($M = 22.0, SD = 1.4$).
Regarding their prior experience with AR systems, one participant indicated, ``I have never used it at all," 13 participants responded, ``I have used it a few times," and two participants selected, ``I use it about once a week."
To assess participants' anxiety levels when talking to new people, a questionnaire was administered. Three participants responded with ``very nervous" or ``nervous," nine participants selected ``somewhat nervous," two participants indicated ``somewhat not nervous," and two participants reported ``not nervous" or ``not nervous at all."
Each participant received a 1,500-yen Amazon gift card as compensation for completing both Experiment 1 and Experiment 2.

\subsection{Conditions}
\label{Conditions}
The two information presentation methods were compared in a within-subject experimental design.
Condition 1 follows the information presentation method and UI of the proposed system, as described in Section 4.
In this condition, the information initially appears over the conversational partner’s face, then moves downward after a set period, and later returns to its original position.
Condition 2 employs a different approach: a fixed text window is displayed in the world coordinate system near the conversational partner’s head and remains stationary throughout the conversation.
This follows the conventional text display method used in HMD-based AR systems.

\subsection{Criteria}
\label{Criteria}
In each condition, the conversation partner evaluated how long the system user read the displayed text.
Specifically, a Visual Analogue Scale (VAS) was used to measure the proportion of time from the start of the interaction during which the system user seemed to be reading, based on observable behaviors, primarily eye movements.
Additionally, the conversation partner rated the appropriateness of the system user's eye movements on a 7-point scale in response to the question: ``If this were a real conversation, how appropriate were the other participant’s eye movements?"

The system users rated the difficulty of reading the text using a seven-point scale as follows:
1: fairly difficult; 2: difficult; 3: somewhat difficult; 4: neither; 5: somewhat easy; 6: easy; 7: fairly easy
Additionally, they were asked to imagine the difficulty of maintaining a natural conversation while occasionally checking the displayed text, assuming they were engaged in a real conversation, and to respond using the same seven-point scale.
Furthermore, the time required to read the text was also measured.

\subsection{Procedures}
First, two participants were called in at a time, provided with an explanation of the experiment outline, and asked to sign an informed consent document.
After receiving instructions on how to use the device (HoloLens2) and software, each participant underwent approximately five minutes of training on basic button operations and viewing information.
Next, the participants sat facing each other on chairs positioned 1.5 meters apart. 
The experimenter explained that the purpose of the experiment was to analyze behavior during conversation.
One participant wore the HoloLens2 as the system user, while the other, serving as the conversational partner, did not wear the device.
The system user pressed a virtual button with their finger, triggering the display of a sentence of approximately 130 characters in Japanese (equivalent to around 200 to 300 characters in English) using one of the presentation methods described in Section \ref{Conditions}.
System users were instructed to read the displayed text as quickly as possible while keeping in mind that they were in the middle of a conversation and to provide a verbal signal upon completion.
 Meanwhile, the conversational partner observed the system user and evaluated their behavior.
This procedure constituted one trial.
The same procedure was then repeated under the alternate display condition, which was switched by pressing the virtual button.
After completing both trials, both participants responded to the questionnaire outlined in Section \ref{Criteria}.
Subsequently, the participants switched roles—those who had been system users became conversational partners and vice versa—and the procedure was repeated.

\subsection{Results}
All data were tested for normality using the Shapiro-Wilk test and for homogeneity of variance using Levene’s test.
If both normality and equal variance were confirmed, ANOVA was applied; otherwise, the Wilcoxon signed-rank test was used.
The significance level was set at 0.05.

\subsubsection{System User Evaluation}
\begin{figure}[t]
 \centering
  \includegraphics[width=1.0 \columnwidth]{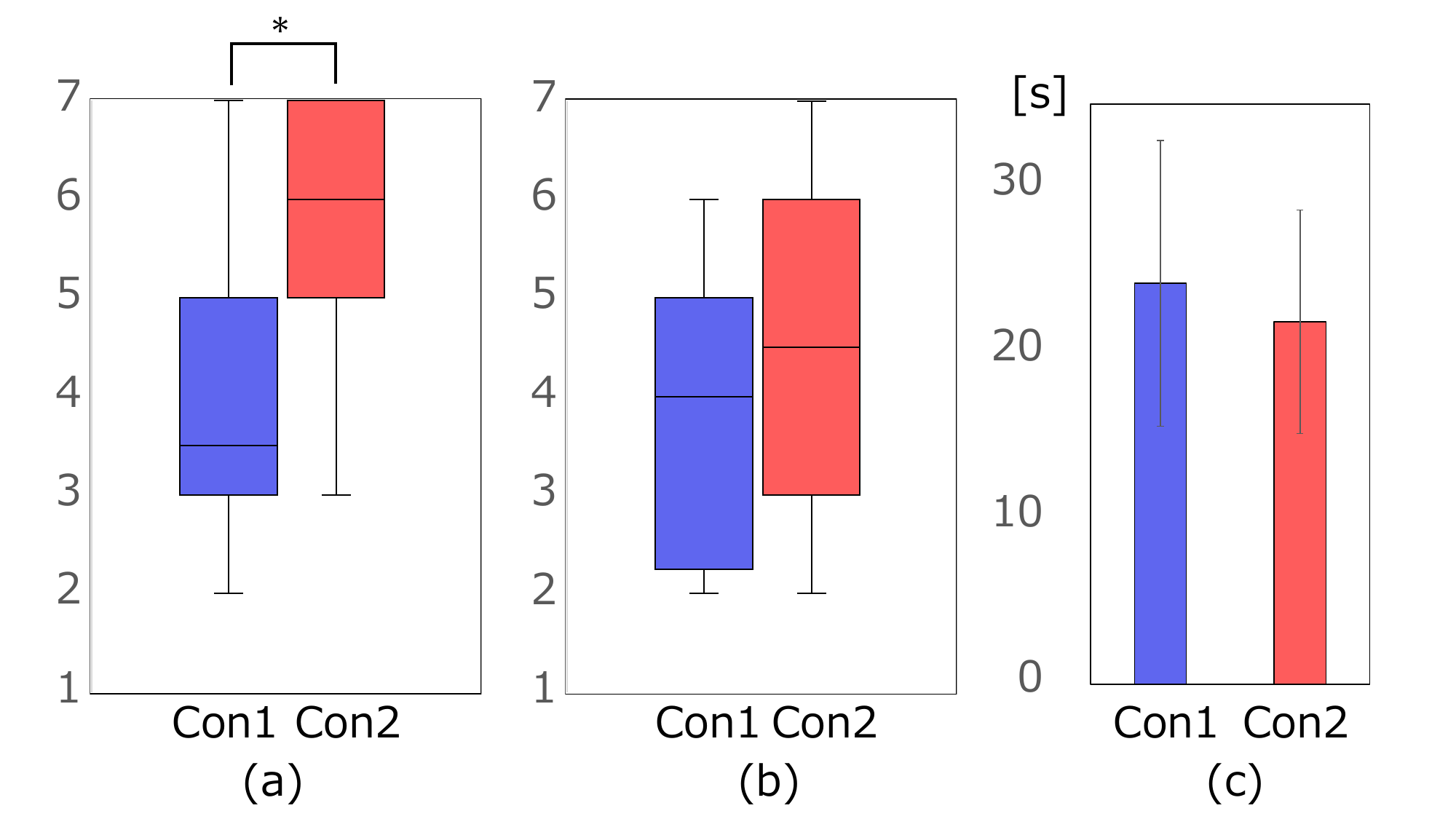}
    \caption{Experiment 1 (Evaluation from System Users): (a) Difficulty in reading the text (1: Difficult, 7: Easy), (b) Difficulty in maintaining a conversation while reading the text (1: Difficult, 7: Easy), (c) Time taken to read the text (seconds).}
 \label{fig:exp1-1}
\end{figure}

This section presents the evaluation from the system user’s perspective.
Figure \ref{fig:exp1-1} (a) illustrates the difficulty of reading the text.
In Condition 1 (proposed), the mean score was 4.0 with a standard deviation (SD) of 1.4, whereas in Condition 2, the mean score was 5.8 ($SD = 1.3$), indicating that reading the text in Condition 1 was significantly more difficult ($p<0.01$).

Figure \ref{fig:exp1-1} (b) illustrates the difficulty of carrying on a conversation while reading a text.
The perceived difficulty of maintaining a conversation while reading showed a mean score of 4.0 ($SD = 1.5$) in Condition 1 and 4.4 ($SD = 1.6$) in Condition 2, with no statistically significant difference between the conditions ($p = 0.449$).

Figure \ref{fig:exp1-1} (c) presents the time required to read the sentence.
The mean reading time in Condition 1 was 24.2 seconds ($SD = 8.6$), compared to 21.9 seconds ($SD = 6.7$) in Condition 2, which also did not show a statistically significant difference ($p = 0.404$).

\subsubsection{Conversation Partner Evaluation}
\begin{figure}[t]
 \centering
  \includegraphics[width=0.9 \columnwidth]{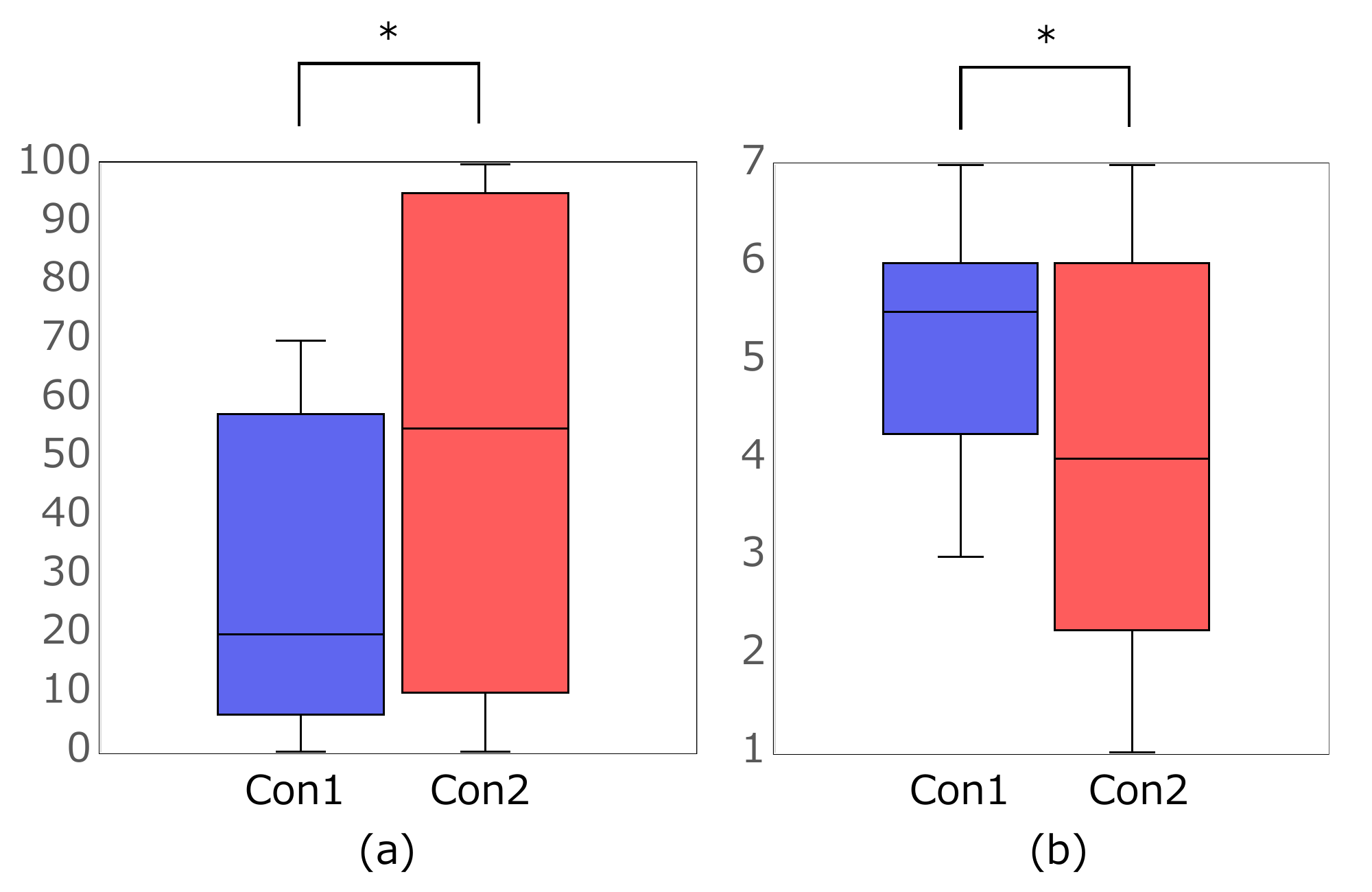}
    \caption{Experiment 1 (Evaluation from Conversation Partners): (a) Percentage of time the system user appeared to be reading the text ($\%$), (b) Appropriateness of the system user's gaze (1: Inappropriate, 7: Appropriate).}
 \label{fig:exp1-2}
\end{figure}

This section presents the evaluation from the perspective of the conversation partner.
Figure \ref{fig:exp1-2}(a) illustrates the percentage of time the system user appeared to be reading the text.
In Condition 1 (proposed), the mean was 29.3$\%$ ($SD = 26.1$), whereas in Condition 2, the mean was 51.2$\%$ ($SD = 40.3$), indicating that Condition 1 resulted in a significantly shorter reading duration ($p=0.012$).

Figure \ref{fig:exp1-2}(b) presents the appropriateness of the system user's gaze (1: Inappropriate, 7: Appropriate).
In Condition 1 (proposed), the mean was 5.3 ($SD = 1.1$), while in Condition 2, the mean was 4.1 ($SD = 2.0$), demonstrating that Condition 1 was perceived as significantly more appropriate ($p=0.027$).

\subsection{Discussion}
Regarding the difficulty of reading the text, Condition 1 (proposed method) was perceived as more challenging.
Two participants noted that the vertical movement of the text made it somewhat difficult to read.
However, from the perspective of the conversation partner, the proportion of time the system user appeared to be reading was significantly lower with the proposed method, and the user's gaze was judged to be more appropriate for a conversation.
Additional feedback on the proposed system included comments such as, ``Even though I observed the user's gaze, it was difficult to perceive that they were reading the text." and ``Apart from occasional up-and-down movements, the user appeared to be focusing almost exclusively on a single point (my eyes)."
These findings suggest that while the proposed method may require some adaptation for reading, it creates a more natural impression for the conversation partner.

There were no significant differences between conditions in terms of reading speed or the difficulty of continuing the conversation while reading the text. 
However, many participants reported experiencing some level of difficulty. Addressing this challenge remains an important area for future improvement.

\section{Experiment2}
\label{experiment2}

Following Experiment 1, Experiment 2 was conducted to evaluate the effectiveness of the proposed system, chatAR, in facilitating conversation.

\subsection{Overview}
Each participant took part individually as a conversation partner, engaging in a two-minute free-form conversation with a system user (the experimenter) wearing the HoloLens 2.
A topic was selected for which the conversation partner had familiarity, while the system user possessed only limited prior knowledge.
After the conversation, participants evaluated their impressions of the interaction using a questionnaire.
To assess the effectiveness of the proposed method, we compared conditions with and without the system’s support.

\subsection{Participants}
The same 16 participants from Experiment 1 took part in Experiment 2.
For all sessions, the system user was a single experimenter who was proficient in operating the proposed system.

\subsection{Conditions}
\label{Conditions2}
The experiment employed a within-subjects design, in which each participant experienced both conditions: using and not using the proposed chatAR system.
To control for viewing conditions, the experimenter wore a HoloLens2 in both conditions.
The chat topics were selected from the following list of topics that the experimenters had no experience with and little knowledge of.
For the two conditions, different topics were chosen by each participant to avoid repetition.

Dialogue topic list are as follows.
Soccer/futsal,
baseball,
golf,
camping,
bouldering,
mountain climbing,
fishing,
muscle training,
surfing,
skiing/snowboarding,
theater-going,
road cycling/cycling,
keeping pets,
K-pop,
favorites (idols, anime, manga, etc.),
art appreciation,
drawing/painting,
wine/sake,
astronomy,
gardening,
cooking,
knitting, handicrafts, making small items,
online games,
smartphone games,
mahjong,
investments/stocks,
horse racing, and
yoga and pilates

\subsection{Criteria}
\label{Criteria2}
The participant evaluated the appropriateness of the system user’s eye movements on a 7-point Likert scale in response to the question: ``how appropriate were the partner's eye movements for conversation?".
The participant also rated the smoothness of the conversation using the question: ``Was the conversation smooth?" and the liveliness of the conversation using: ``Was the conversation lively?" Both were assessed on the same 7-point scale.
Each score meant, ``1: strongly disagree. 2: disagree. 3: somewhat disagree. 4: neither agree nor disagree. 5: somewhat agree. 6: agree. 7: strongly agree".

We also analyzed the amount of speech and the number of turn-takings by the participants and the system user (experimenter) during the conversation.
For speech analysis, the recorded audio data from the conversation was transcribed using Azure Speech to Text \cite{SpeechServices}. The transcription was separated by speaker and then converted into hiragana, the smallest phonetic unit in the Japanese language.
Based on this processed data, the number of turn-takings for each participant was calculated. 
Note that utterances consisting solely of filler phrases or simple repetition of the conversational partner’s previous words were not counted as turn-takings.

\subsection{Procedures}
After completing the first experiment, participants took a five-minute break.
Then, one participant was invited into a separate room. 
Each participant was asked to select two topics of interest from the list provided in Section \ref{Conditions}.
The participant and the experimenter then sat facing each other on chairs placed 1.5 meters apart.
Participants were instructed to engage in a natural conversation, assuming a scenario such as ``chatting with someone you have just met at an on-campus or off-campus social event.".
The system user (experimenter) wore a HoloLens2.
When the start button was pressed, information was presented either using the proposed method or not, depending on the condition assigned.
Participants initiated the conversation by stating, ``I like ***** (e.g., Soccer)" referring to one of the topics they had selected earlier. 
They then engaged in free conversation for two minutes.
After two minutes, a timer sound was automatically played to indicate the end of the conversation, even if someone was in the middle of speaking.
Subsequently, when the start button was pressed again, information was presented under the alternate experimental condition.
The order of the two conditions was counterbalanced across participants to ensure equal distribution.
Participants then had another two-minute conversation on the second selected topic, following the same procedure.
In both conditions, the system user primarily focused on listening but also attempted to facilitate the conversation as appropriate.
The difference between the two conditions was not disclosed to the participants.
After both conversations, participants completed a questionnaire evaluating their impressions regarding the liveliness of the conversation and the behavior of the system user.

\subsection{Results}
\begin{figure*}[t]
 \centering
  \includegraphics[width=1.8 \columnwidth]{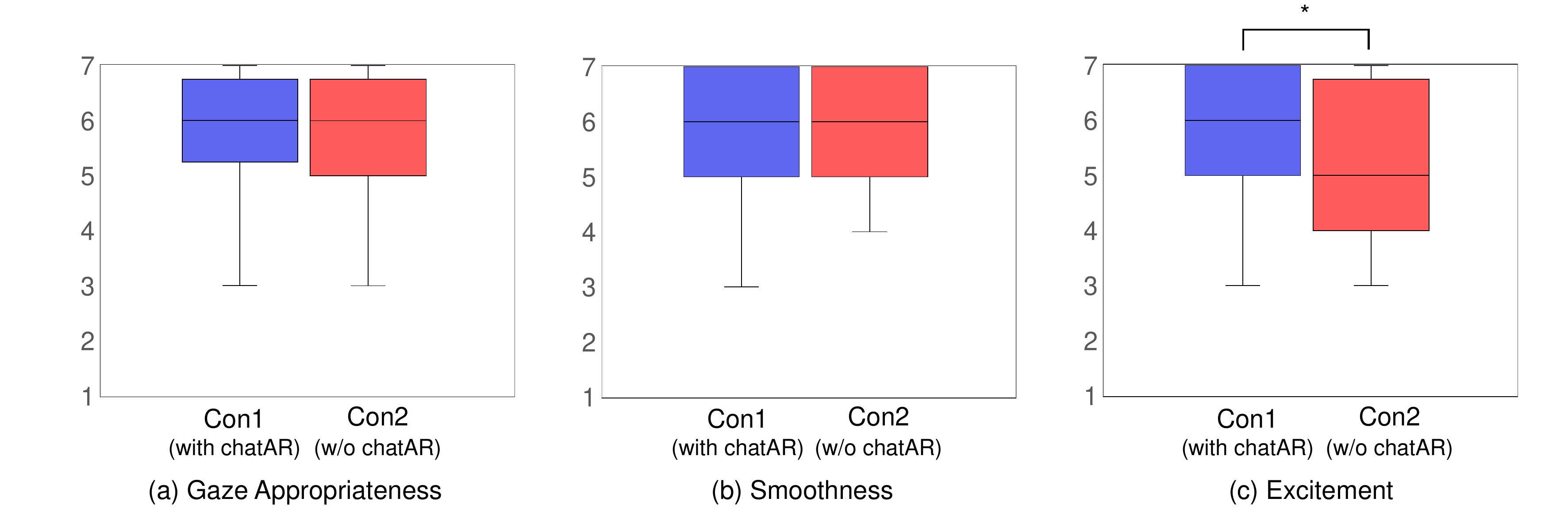}
    \caption{Experiment 2: (a) Appropriateness of gaze (1 = inappropriate, 7 = appropriate), (b) Smoothness of conversation (1 = not smooth, 7 = smooth), (c) Excitement of conversation (1 = not exciting, 7 = exciting)}
 \label{fig:exp2}
\end{figure*}

The statistical analysis methods were the same as those used in Experiment 1.
Figure \ref{fig:exp2} (a) presents the ratings of the appropriateness of the conversation partner's gaze (1: inappropriate, 7: appropriate).
In Condition 1 (with chatAR), the mean rating was 5.7 ($SD = 1.3$), and in Condition 2 (without chatAR), the mean was 5.6 ($SD = 1.2$), showing no significant difference ($p = 0.762$).

Figure \ref{fig:exp2} (b) shows the ratings for the smoothness of the conversation (1: not smooth, 7: smooth).
The mean rating for Condition 1 was 5.8 ($SD = 1.3$), and for Condition 2, 5.9 ($SD = 1.1$), also showing no significant difference ($p = 0.809$).

Figure \ref{fig:exp2} (c) shows the participants’ ratings of how exciting the conversation was (1: not exciting, 7: very exciting).
In Condition 1 (with chatAR), the mean rating was 5.8 ($SD = 1.1$), while in Condition 2 (without chatAR), the mean was 5.1 ($SD = 1.3$), indicating that participants felt significantly more excitement when using the proposed system ($p = 0.046$).

\begin{figure}[t]
 \centering
  \includegraphics[width=1.0 \columnwidth]{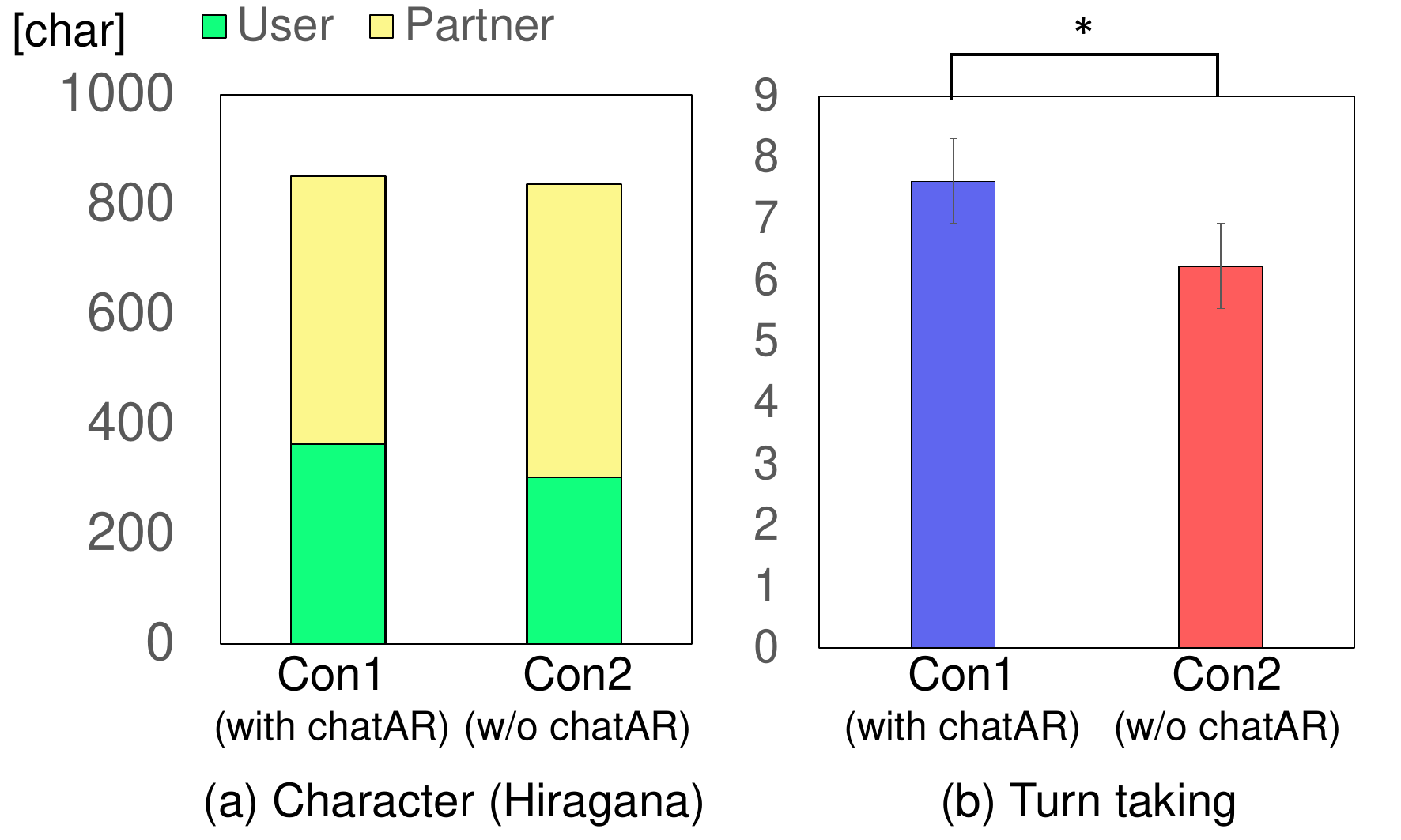}
    \caption{Experiment 2: (a) Number of characters (in hiragana) uttered by the system user and the conversation partner, (b) Number of turn-takings by the conversation partner (the system user).}
 \label{fig:exp2-2}
\end{figure}

Figure \ref{fig:exp2-2} (a) presents the average number of characters (hiragana) spoken by each speaker.
While the total number of utterances was similar between the two conditions, the proportion of utterances made by the system user increased in Condition 1 compared to Condition 2.
Figure \ref{fig:exp2-2} (b) shows the number of turn-takings by the participant (conversation partner).
In Condition 1, the mean was 7.6 ($SD = 1.5$), and in Condition 2, the mean was 6.2 ($SD = 0.9$), with Condition 1 showing significantly more turn-takings ($p = 0.024$).

\begin{figure*}[t]
 \centering
  \includegraphics[width=2.0 \columnwidth]{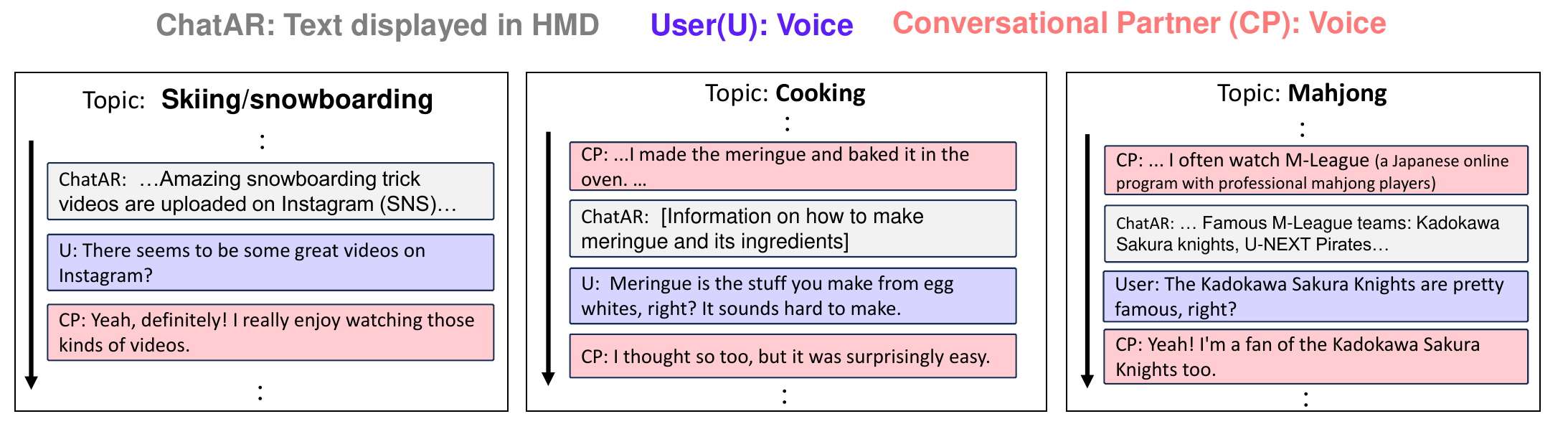}
    \caption{Experiment 2: Examples of support provided by chatAR (excerpts from actual conversations). All conversations were originally conducted in Japanese and have been translated into English for presentation purposes.}
 \label{fig:example-of-conv}
\end{figure*}

Figure \ref{fig:example-of-conv} provides examples of conversational support provided by chatAR.

\subsection{Discussion}
There was no significant difference in participants’ ratings of the appropriateness of their partner’s gaze between Condition 1 (with chatAR) and Condition 2 (without chatAR).
This suggests that, at least when the user is familiar with chatAR, the use of the system does not significantly disrupt gaze behavior during conversation.

While the smoothness of the conversation was similar across both conditions, the level of excitement was slightly higher when the system was used.
With chatAR, the amount of the user's speech increased and the number of turn-taking sessions between the two also increased.
This indicates that the system user was able to provide more informative responses rather than merely giving backchannel feedback, which may have contributed to the heightened conversational excitement.
This interpretation is supported by a participant’s comment: ``I was able to talk about what I wanted because my conversation partner (the system user) guided the discussion toward topics that were relevant to me."

Most participants did not seem to recognize the differences in system behavior between the two conditions.
This suggests that the use of the system did not significantly interfere with the flow of conversation.
However, one participant expressed dissatisfaction with the system usage condition, stating that he still had something he wanted to talk about when the topic suddenly changed.
This disruption was caused by the experimenter introducing newly displayed relevant information, prompting a shift in topic.
Mitigating such negative effects remains a challenge for future work.

\section{General Discussion}

\subsection{Ethical Considerations On System and Concepts}
The two experiments showed that the proposed system enhances the excitement of the conversation while maintaining the user's gaze, which is favorable to the conversational partner.
This suggests that the proposed system offers a degree of effectiveness in providing conversational support and could serve as a promising solution in the future for individuals who feel uncomfortable or anxious in social conversational situations.

However, the ethical or mannerly acceptability of accessing information beyond one's own knowledge without informing one's conversation partner is crucial for the global adoption of such a system.
This problem may be resolved in the future if such systems become socially accepted and widespread, but until then, it may be necessary to take precautions such as obtaining permission from the other party in advance.

In addition, there is concern that when the system is used continuously, the conversational partner may develop an image of the system user that differs from reality in terms of knowledge and values, and resolving this issue is also an important challenge.

In the current system design, we presented only a limited number of pieces of information related to the chat topic and left the decision of whether and how to use this information to the user.
However, a more proactive method of support that relies on AI to determine the content of conversations is also possible.
In a more proactive system, ethical considerations regarding the influence of generative AI on the content of human conversations are crucial.
It is also undesirable for the use of the system to undermine the user's communication and conceptual abilities.
Designs that avoid such side effects should be continuously considered.

\subsection{Limitation and Future Work}
The current system lacks optimal timing for automatic keyword extraction and information generation, resulting in updates that may occur at undesirable moments for users.
The current mitigation measure, a gaze-operated generation stop/resume button, is suboptimal due to the high cognitive load caused by conscious eye movement.
In relation to this, the current system has a delay of about 2 seconds between the utterance of keywords and the generation of information, so users will need to practice using it in order to be able to continue a natural conversation.
Therefore, it is a future challenge to make the timing of information generation more suitable by considering the conversational context.

In addition, due to the control of conversation topics and the difficulties in using the current system described above, the system was used only by the experimenter in Experiment 2.
The next step is to have participants use the updated system and evaluate its effectiveness in providing conversational support.

\section{Conclusion}

In this study, we proposed ChatAR, a support system combining AR technology and LLM to facilitate interpersonal interaction.
Experiment 1 demonstrated that the proposed information presentation method aligns better with the user's eye movement toward the conversation partner compared to the method of presenting information at a fixed position.
Experiment 2 demonstrated that the proposed system contributes to enhancing the liveliness of conversations.
On the other hand, there were a few instances where the timing of the supplementary information presentation inadvertently interfered with the conversation.
In the future, we plan to automatically identify appropriate keywords by taking into account the conversational context and non-verbal cues of the user and conversational partner. 
Additionally, we aim to validate the system's effectiveness through participant usage.

\acknowledgments{
This work was supported by JSPS KAKENHI Grant Number 25K01204, Japan.
}

\bibliographystyle{abbrv-doi}


\end{document}